\documentclass[conference,a4paper]{IEEEtran}
\IEEEoverridecommandlockouts
\usepackage{enumerate}
\usepackage[english]{babel}
\usepackage[utf8]{inputenc}
\usepackage{graphicx}  
\usepackage{amsmath}
\usepackage{booktabs}
\usepackage{multirow}
\usepackage{subcaption}
\usepackage{cite}
\usepackage{amsmath}
\usepackage{amssymb}
\usepackage{graphicx}
\usepackage{hyperref}
\usepackage{graphics}
\usepackage[linesnumbered,ruled]{algorithm2e}
\newtheorem{remark}{Remark}

\usepackage{upgreek}

\usepackage{mathtools}

\usepackage{textcomp}
\usepackage{hyperref}
\usepackage{lipsum}
\usepackage{tikz}

\newcommand\copyrighttext{%
  \footnotesize A version of this paper has been accepted at 2020 IEEE PES Innovative Smart Grid Technologies Europe (ISGT-Europe). Details of the published paper are as follows: Title: Stochastic Game Frameworks for Efficient Energy Management in Microgrid Networks. DOI:  10.1109/ISGT-Europe47291.2020.9248952. \textcopyright 2020 IEEE. Personal use is permitted, but republication/redistribution requires IEEE permission. }
\newcommand\copyrightnotice{%
\begin{tikzpicture}[remember picture,overlay]
\node[anchor=south,yshift=10pt] at (current page.south) {\fbox{\parbox{\dimexpr\textwidth-\fboxsep-\fboxrule\relax}{\copyrighttext}}};
\end{tikzpicture}%
}

\ifCLASSINFOpdf
 
\else
  
\fi

\title{A Stochastic Game Framework for Efficient Energy
Management in Microgrid Networks}  

\author{Shravan	Nayak$^{1}$, Chanakya Ajit Ekbote$^{2}$, Annanya Pratap Singh Chauhan$^{3}$, Raghuram Bharadwaj Diddigi$^{4}$, \\ Prishita Ray$^{5}$, Abhinava Sikdar$^{6}$, Sai Koti Reddy Danda$^{7}$, Shalabh Bhatnagar$^{8}$
\thanks{$^{*}$ Equal contribution by the first three authors}
\thanks{$^{1}$ IIT Varanasi, India. 
        {\tt\small pshravan.nayak.ece17@itbhu.ac.in}}%
\thanks{$^{2}$ IIT Bhubaneswar, India.
        {\tt\small ca10@iitbbs.ac.in}}%
\thanks{$^{3}$ IIT Guwahati, India.
        {\tt\small annan170101008@iitg.ac.in}}
\thanks{$^{4}$ Indian Institute of Science, India.
        {\tt\small raghub@iisc.ac.in}}%
\thanks{$^{5}$ VIT Vellore, India.
        {\tt\small prishita.ray2017@vitstudent.ac.in}}%
\thanks{$^{6}$ IIT Delhi, India.
        {\tt\small asikdar.iitd@gmail.com}}%
\thanks{$^{7}$ IBM Research - India.
        {\tt\small d.saikotireddy@in.ibm.com}}%
\thanks{$^{8}$ Indian Institute of Science, India.
         {\tt\small shalabh@iisc.ac.in}}
}
\begin{document}
\pagestyle{empty} 
\maketitle
\copyrightnotice
\thispagestyle{empty}

\begin{abstract}  
We consider the problem of energy management in microgrid networks. A microgrid is capable of generating a limited amount of energy from a renewable resource and is responsible for handling the demands of its dedicated customers. Owing to the variable nature of renewable generation and the demands of the customers, it becomes imperative that each microgrid optimally manages its energy. This involves intelligently scheduling the demands at the customer side, selling (when there is a surplus) and buying (when there is a deficit) the power from its neighboring microgrids depending on its current and future needs. Typically, the transaction of power among the microgrids happens at a pre-decided price by the central grid. In this work, we formulate the problems of demand and battery scheduling, energy trading and dynamic pricing (where we allow the microgrids to decide the price of the transaction depending on their current configuration of demand and renewable energy) in the framework of stochastic games. Subsequently, we propose a novel approach that makes use of independent learners Deep Q-learning algorithm to solve this problem. Through extensive empirical evaluation, we show that our proposed framework is more beneficial to the majority of the microgrids and we provide a detailed analysis of the results.  
\end{abstract}


\section{Introduction}
Microgrid networks are a collection of small-scale renewable energy resources that fulfil local consumer demands. They may function independently or in collaboration with other microgrids by trading energy. The main advantage of using microgrids is their ability to decentralize power distribution from the central grid, thus providing a more efficient architecture for energy distribution by targeting smaller areas and serving as reliable power sources when the central grid has a deficiency. In addition, they also reduce losses incurred due to long-distance energy transmissions and prove to be a more cost-effective and eco-friendly alternative to traditional resources such as fossil fuels that cause more pollution and are depleting at an alarming rate. Their main tasks include local power generation, storing energy, trading power with other microgrids and satiating local consumer demands.

At the demand side, customers have certain flexible demands that can be satisfied any time during certain given time periods throughout the day. These loads are structured in such a way that they can be fulfilled anytime during their allotted time period. For example: If a washing machine is being operated any time between 2 pm to 6 pm at a particular household, the microgrid would have the ability to intelligently provide the energy required to run the washing machine at any time during this period. These demands are classified as Activities of Daily Living (ADL). Each microgrid has the ability to schedule these ADL demands depending on the peak demand as well as the local energy generation. ADL scheduling does not reduce power consumption; it merely helps in reducing the peak load at any time instance.


Energy trading plays a vital role in the decentralization of power generation and maintaining stability at the microgrid sites. This involves buying and selling the power among neighboring microgrids at favorable prices. The main focus of our paper is to highlight advantages of using the dynamic pricing scheme (a scheme that allows the microgrids to select the prices at which it decides to sell power) in tandem with ADL scheduling. The dynamic pricing scheme not only provides the microgrids autonomy for selecting prices according to their convenience (based on their current state), but also encourages energy trading amongst them. This promotes more cooperation amongst microgrids thereby causing a lower dependency on the central grid for fulfilling local energy requirements. This in turn enables better decentralization, and also helps individual microgrids obtain higher rewards as our results show, than if they were to follow a constant pricing policy. ADL scheduling not only helps reduce the peak load, it also allows the microgrid to intelligently defer certain loads while selling the remaining energy in order to optimize the overall reward it receives. 

The literature on the energy trading between microgrid networks is vast. The problem of energy trading in microgrid networks has been primarily considered from three different points of view. In \cite{lee2015distributed,wang2016incentivizing,li2017risk}, game theoretic models have been proposed along with the equilibrium analysis of solutions. In \cite{chaouachi2012multiobjective,nunna2013energy,gregoratti2014distributed,shi2014distributed,liu2018dynamic}, the energy trading has been formulated as an optimization problem and models such as convex programming and Linear programming have been used to compute optimal solutions. The third popular framework for energy trading in microgrids is Reinforcement Learning (RL). 
RL is a popular paradigm that provides learning algorithms for computing the solution when the model information is not known. We now discuss some of the works that propose RL algorithms to solve the energy trading problem among microgrids. In \cite{xiao2017energy}, an energy trading game using RL techniques has been proposed. In their model, each microgrid, based on its current state configurations, computes the amount of energy to be traded with neighboring microgrids in order to maximize its rewards. However, the prices in this model are the market prices and are not dynamic. In \cite{kim2015dynamic,lu2018dynamic}, a dynamic pricing problem for a single microgrid is considered. Based on the consumption pattern of its customers, the microgrid decides the price of the power to be sold to its customers. In \cite{wang2016reinforcement}, a novel energy trading model for microgrid networks is proposed that considers dynamic pricing. However, the dynamic scheduling of customers demand is not considered. Deep Reinforcement Learning algorithms have been successfully applied for computing optimal solutions in the context of energy trading between microgrids in \cite{xiao2018reinforcement,chen2018local}, for storage device management in \cite{franccois2016deep}, and for energy management in \cite{lu2019incentive,ji2019real}. The closest work to ours is \cite{diddigi2017unified}, where an energy trading model for a microgrid network has been proposed that also considers job scheduling for customers. We extend this model considerably to include dynamic pricing for transactions between microgrids and apply the independent learners Deep Q-learning algorithm that is shown to have a good empirical performance in literature \cite{tampuu2017multiagent}.

In \cite{vazquez2019reinforcement}, an extensive survey of RL algorithms for demand response is carried out. They also identify the need for RL algorithms to consider demand response in multi-agent scenarios with demand-dependent dynamic prices. Our work is a step in this direction. 

Our main contributions in this paper are as follows:

\begin{enumerate}
    \item We construct a Multi-Agent Reinforcement Learning framework that addresses the supply-side management problems of dynamic pricing, battery scheduling as well as the demand-side management problem of scheduling ADL jobs.
    \item To the best of our knowledge, ours is the first work that uses a novel DQN approach to solve both these problems by creating two separate neural networks (for handling the tasks of stochastic job scheduling as well as energy trading) both working as ingredients to the same Markov Decision Process.
    \item We perform experiments to show that the reward obtained by a microgrid is lower if it employs a constant pricing policy instead of a dynamic pricing policy as the latter ensures better participation in energy trading.
    \item  We empirically show that the proposed dynamic pricing setup ensures more reward for most of the participating microgrids.
    \item Based on the results of our experiments we also provide detailed analysis on the behaviour of microgrids under various setups.
\end{enumerate}

\section{Problem Formulation}
In this section, we describe the model of the microgrids that enables the energy trading and job scheduling. Our solution is based on a framework of the problem that consists of independent microgrids, interconnected by multiple transmission lines, in the presence of a central grid. Each microgrid has the ability to locally generate renewable energy and it also has the provision of storing energy in a battery unit.
 We divide each day into several time steps of equal duration, for better granularity of the decision-making process. At each time step, the microgrids have information about their current local demand, the renewable energy generated, the amount of energy stored in the battery as well as the remaining ADL demands that are to be fulfilled in that day. Depending on this information the microgrids make decisions regarding their demand and supply management at regular time intervals. 
These decisions are as follows:
\begin{itemize}
    \item The scheduling and fulfillment of ADL demands.
    \item The fulfillment of non-ADL demands.
    \item The amount of electricity to buy or sell, and also the price at which to sell electricity.
    \item The amount of energy to be stored in the battery.
\end{itemize}

We formulate this problem in the framework of stochastic games. A stochastic game is a popular framework that is used for modeling competing or cooperative agents in a stochastic environment \cite{bowling2000analysis}. The main ingredient of a stochastic game is the tuple $<n,S,A,P,r_1,r_2,\ldots,r_n,\gamma>$, where $n$ is the number of agents, $S = S_1 \times S_2 \times \ldots \times S_n$ denotes the joint state space where $S_i$ is the state space of agent $i$, $A = A_1 \times A_2 \times \ldots \times A_n$ denotes the joint action space with $A_i$ representing the action space of the agent $i$. $P(s'|s,a)$ is the probability transition rule that gives the probability of moving to next state $s' \in S$ when action tuple $a \in A$ is taken in state $s \in S$. Note that in our model, an agent $i$ can only observe its own state $s_i \in S_i$ and picks an action $a_i \in A_i$. Finally, $r_i(s,a)$ is the single-stage reward function of the agent $i$ that gives the reward value obtained when the joint action tuple $a \in A$ is taken in state $s \in S$ and $\gamma$ is the discount factor. The objective of the agent $i$ is to compute a policy a $\pi^{*}_i: S_i \xrightarrow{} A_i$, that maximizes its total discounted reward, given the optimal policies of other agents $\pi^{*}_{-i}$. That is, 
\begin{align}
    \pi_i^* = \arg \max_{\pi_i \in \Pi_i } E\Big{[}\sum_{i=0}^{\infty}\gamma^i r_i(s_i,(\pi_i(s_i),\pi^{*}_{-i}(s_i))\Big{]},
\end{align}
where $\Pi_i$ is set of all policies of agent $i$ and $E(.)$ is the expectation over states $s_i,~ i = 0,\ldots,\infty$ with the initial state $s_0$ sampled from a known distribution $\rho$.

We now describe in detail the states, actions and single-stage rewards of each of the microgrids. 

\subsection{States}
The state of the microgrid $i$ at time $t$ is given by: $S=(t,ne_{t}^{i},d_{t}^{i},ADL_{t}^{i},GP_{t}^{i})$, where:
\begin{enumerate}
    \item Time state ($t$): This is the time interval of the day at which the decision is taken by the microgrid.
    \item The Net Energy ($ne_{t}^{i}$): This is a cumulative sum of battery value ($b_{t}^{i}$) and generated renewable energy ($re_{t}^{i}$) subtracted by the Non-ADL demand ($d_{t}^{i}$). Thus  $ne_{t}^{i}=re_{t}^{i} +b_{t}^{i}-d_{t}^{i}$
    \item Non-ADL Demand ($d_{t}^{i}$) : This signifies the local consumer demand pertaining to that microgrid. This is provided in addition to the net demand so that the agent is able to estimate the cumulative sum of the battery and the renewable generation at that time step. 
    \item ADL state ($ADL_{t}^{i}$): This state component jointly signifies the ADL loads that are remaining and the adl loads that have been completed by the adl agent, till the current time step. 
    \item Grid Price ($GP_{t}^{i}$): This is the price at which a microgrid buys power from the central grid at that time step. \footnote[1]{Please note that when the microgrids sell power to the central grid, the selling price would be $GP_{t}^{i}$ - \textit{k}, where '\textit{k}' is a positive integer. The reason for this will be made clear in the `Design Constraints' section. }

\end{enumerate}
\subsection{Actions}
As discussed above, the actions of the agents involve deciding the ADL demands to be scheduled and the amount of energy to be sold/bought. Additionally, we integrate the pricing model where the microgrids also decide the price at which energy trading takes place. In particular, the actions of the microgrid are as follows:
\begin{itemize}
    \item ADL action ($adl_{t}^{i}$):  This signifies the ADL demands that a microgrid plans to fulfil in the current time step. 
    \item Electricity to be traded ($u_{t}^{i}$): This denotes the amount of power that the microgrid decides to trade amongst other microgrids as well as the central grid. It is governed by a set of constraints that are derived from the net demand, the Non-ADL demand, the ADL action and the battery capacity to ensure that the microgrid remains stable. A negative value of $u_{t}^{i}$ signifies that the microgrid is buying electricity whereas a positive value of $u_{t}^{i}$ signifies that the microgrid is selling electricity.
    \item Price Chosen ($p_{t}^{i}$): This signifies the price chosen by the microgrid at which power is sold. Sellers quote a price while buyers are assumed to adhere to the price determined by the sellers. 
\end{itemize}

After the microgrids select their respective $p_{t}^{i}$ and $u_{t}^{i}$ actions, they are divided into two groups namely buyer microgrids and seller microgrids. A grid is classified as a buyer microgrid based on whether or not the value of $u_{t}^{i}$ selected is negative. Conversely, a grid is classified as a seller microgrid if the value of $u_{t}^{i}$ selected is positive.

Once the microgrids are divided into groups of buyers and sellers, energy trading happens in the following way. First, a microgrid from the seller group (let’s call it the leader), that quotes the lowest price is selected. The amount of energy that the leader microgrid is willing to sell is shared amongst the buyer microgrids, proportional to the energy they demand. This is to ensure that there is no bias amongst buyer microgrids. Once the leader microgrid has sold all of its energy, a new leader is chosen, i.e., the one quoting the next best price. This chain continues on till there are no seller microgrids or no buyer microgrids left in the process.

Even after these transactions, if certain demands of the buyer microgrids remain unfulfilled, the remaining amount of energy is bought from the central grid at a price: $GP_{t}^{i}$. Conversely if all the demands of the buyer microgrids are satiated, the seller microgrids end up selling the remaining $u_{t}^{i}$ to the central grid, at a price: $GP_{t}^{i}$ - \textit{k}. 
\subsection{Reward Function}
The goal of each microgrid is to obtain adequate profits acquired by selling electricity, while satiating local consumer demand which consists of both ADL as well as Non-ADL demand. The reward computed \footnote[2]{Please note that the reward function does not take into account the profits obtained by selling electricity to the local customers.}, takes care of both of these conditions by giving a positive reinforcement to the agents when electricity is sold, charges the microgrid if electricity is bought, and also penalises when the instantaneous local consumer demand (Non-ADL demand)  as well as the ADL demand is not met with.
\begin{align*}
    r_t^{i} = u_{t}^{i} * p_{t}^{i} - k1*(\text{unfulfilled Non-ADL demand})  \\ \nonumber - k1*(\text{unfulfilled ADL demand}),
\end{align*}
where $k1$ is a positive constant. Changing the values of $k1$ leads to the the microgrids exhibiting different behaviors. When $u_t^i$ is much larger than $k1$, the microgrids favour selling energy as compared to satisfying their local consumer demands (both Non-ADL as well as ADL demand). Conversely, when k1 is much larger than $u_t^i$, the microgrids prefer to satiate their local consumer demands (both Non-ADL as well as ADL) as compared to selling energy. This can be explained as follows: each microgrid is tasked with optimising its reward function. By changing $k1$, the weights given to selling energy and satisfying local consumer demand changes, which in turn changes the reward function. To emulate a real world scenario, we have given a higher importance to satisfying local consumer demand as compared to selling energy. In our experiments, $k1$ is set to 30. 

If $u_t^i$ is positive, it implies that the microgrid is selling electricity and hence receives a profit.
If $u_t^i$  is negative, it implies that the microgrid is buying energy and hence it incurs a cost. As $k1$ is a positive constant, the microgrid is penalised for not satisfying the Non-ADL as well as the ADL demand. If the local consumer demands are met with, the microgrid receives no penalty.

Note that the reward of each microgrid depends not only on its own action, but also on the action of other microgrids (as the energy being traded by other microgrids as well as the price they quote implicitly affects the reward of that microgrid) and hence, this structure induces a stochastic game amongst the microgrids.


\section{Proposed Algorithm}
To fulfill the demand-side management tasks as well as supply-side tasks, each microgrid employs two agents. The first agent (also called the ADL agent) is responsible for the demand-side management. It decides which ADL tasks would be scheduled in the current time step, and this information is then provided to the second agent. The second agent (also called the Energy Trading (ET) agent) is responsible for the supply-side management. It decides the units of electricity to buy or sell, and also sets the transaction prices, i.e., the prices at which the energy trading happens.

\begin{figure*}{}
    \centering
    \includegraphics[scale = 0.7]{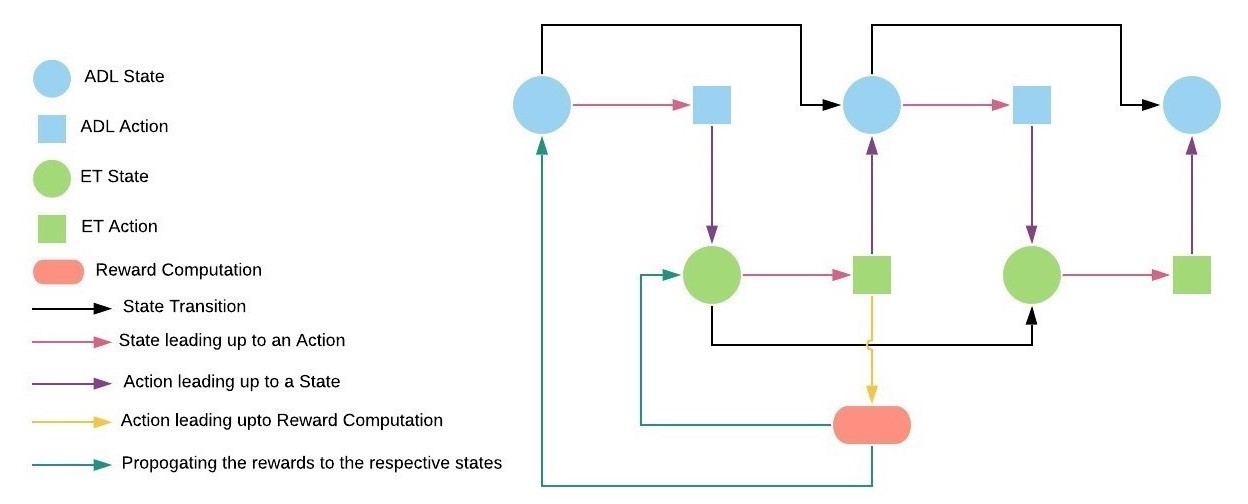}
    \caption{Interplay of states and actions between ADL agent and ET agent}
    \label{f4}
\end{figure*}

Based on the actions taken by the ADL and ET agents, a common reward is obtained by both the agents. This can be justified by the reasoning that both the agents are cohesively working in order to fulfill a common higher goal. Hence, the same credit will be assigned to both of these agents.
Due to the interplay between the ADL and ET agents, a single MDP is created which models the state transitions, action selection as well as the reward computation for both the agents. This interplay is shown in Figure \ref{f4}. 

The advantages of using the two separate networks that share the same rewards are as follows: (a). By creating two networks that perform two different tasks that help fulfill a common goal, we have devised a method to successfully model the execution of sequential tasks, using RL. Moreover, by propagating the same reward to both the networks, we have also empirically shown that sharing the same reward for modeling sequential tasks does lead to network learning. Such kind of a sequential learning approach can be used for a lot of real-world fields such as robotics, auctions etc (b). By creating two networks, instead of one very large network, we reduce the number of iterations needed to obtain optimal policy as this enables better exploration of the action space (c). The interplay between the networks as shown in the paper is also novel to the best of our knowledge. 

Note that both, the ADL Agent and the ET agent have the same state space except for one parameter. The ADL agent has a parameter known as the ADL state (which signifies which ADL actions have to be fulfilled). Instead of this parameter, the ET agent has a parameter known as ADL action ( the action chosen by the ADL agent). Hence the replay buffers for both the agents are similar. Therefore, by sharing a similar state space and reward, the agents are cooperating. Moreover since the ADL and ET agents have to optimize (increase) their rewards, they would implicitly cooperate to obtain an optimal policy.

To optimise the long term discounted rewards obtained by each microgrid, each agent uses the Deep Q-learning algorithm \cite{mnih2013playing}. This is further described in detail in Algorithm 1.

\begin{algorithm}[h]
    Initialize the ADL network with random weights $\lambda$. \\
    Initialize the ET network with random weights $\theta$. \\
    Initialize the replay memory ($M$) to capacity $D$.\\
    \For{\text{episode 1 to N}}{
    Take an ADL action ($adl_{t}^{i}$) using an $\epsilon-$greedy policy via the ADL Network ($Q_{adl}$). \\
    Using  the chosen $adl_{t}^{i}$, select actions $p_{t}^{i}$, $u_{t}^{i}$ using an $\epsilon-$greedy policy via the ET Network ($Q_{et}$). \\
    Observe the rewards ($r_{t}^{i}$) and next states ($s'_{adl}, s'_{et}$) obtained by executing the actions. \\
    Store tuple $(s_{adl},s_{et},adl,p_{t}^{i},u_{t}^{i},r_{t}^{i},s'_{adl},s'_{et})$ in $M$. \\
        Sample a mini batch of $(s_{adl},s_{et},adl_{t}^{i},p_{t}^{i},u_{t}^{i},r_{t}^{i},s'_{adl},s'_{et})$ from $M$ and set : \newline
         $\phi=r_{t}^{i}+\gamma* \max_{a}(Q_{adl}(s_{adl}^{'},a |\lambda))$ 
         $\psi=r_{t}^{i}+\gamma*  \max_{p,u}(Q_{et}(s_{et}^{'},p,u |\theta))$\\
         Perform the gradient descent step for the ADL network on the loss function given by: $(\phi-Q_{adl}(s_{adl},adl_{t}^{i}|\lambda))^{2}$ \\
         Perform the gradient descent step for the ET network on the loss function given by: $(\psi-Q_{et}(s_{et},p_{t}^{i}, u_{t}^{i}|\theta))^{2}$

    }
	\label{Algorithm}
	\caption{Proposed Algorithm}
\end{algorithm}

\section{Design Constraints}

In this section, we describe the constraints that we impose on the proposed model. 
\subsection{Price Constraints}

To ensure that transactions occur between microgrids, the microgrids are allowed to sell energy within a price range of $[gp - k,\; gp]$ (where $gp$ is the central grid price and $k$ is a positive constant). This can be justified as follows. If a microgrid quotes a price higher than that quoted by the central grid, the transactions would not even occur as the other microgrids would prefer to buy directly from the central grid.

Moreover, to ensure that energy trading occurs between microgrids, they are allowed to sell power to the central grid at the least price the microgrid can quote, i.e., $gp - k$. This would ensure that a microgrid would prefer to sell to another microgrid as compared to the central grid. Next, we describe the constraints, that are required to maintain stability at the microgrids. 
\subsection{Energy Trading Constraints}
For emulating a real world scenario, it becomes imperative that real world constraints are imposed on the amount of energy bought or sold. These constraints are dependant on physical limitations such as the maximum battery capacity, the max energy that can be handled by each microgrid etc. The constraints are imposed as follows: \\ \\
\textbf{a. Lower bound on the amount of electricity traded:}
\begin{align}\label{lbc}
   u_t^i \geq max(-M,\:ne_{t}^{ i}-F(A_{t}^{ i}) -B).
\end{align}

The first term $-M$ depicts that a microgrid cannot be allowed to buy more than $M$ amount of electricity, thus preventing the microgrid circuits from being excessively overloaded due to the inflow of excess energy. Thus $-M \leq u_{t}^{i}$.

After each transaction, the amount of energy that would be stored in the battery of each microgrid, (after factoring in the energy generated, the Non-ADL demand, the ADL demand, the ADL action selected and the energy present in the battery prior to the transaction) would be less than or equal to the the maximum battery capacity, hence preventing the microgrids from buying excess energy and then in turn, wasting it. Thus, 
\begin{align}
    & ne_{t}^{i}-F(A_{t}^{i}) -u_{t}^i \leq B, ~ \text{which implies} \\
    &ne_{t}^{i}-F(A_{t}^{i}) - B \leq u_{t}^i,
\end{align}
where $F(A_{t}^{i})$ represents the units of energy that are required to fulfill the selected ADL action.

The second term in the max function in \eqref{lbc} ensures that the ADL action selected by the ADL Network is fulfilled. 

A maximum of the above two terms is taken to allow the microgrid to trade the maximum energy possible whilst fulfilling the decided ADL actions and also taking the microgrid stability into consideration as well.\\


\textbf{b. Upper bound on the amount of electricity traded:}
\begin{align}
     u_t^i \leq ne_{t}^{ i} + d_{t}^{i} -F(A_{t}^{ i}).
\end{align}

The upper bound is derived from the fact that once an ADL action has been chosen then it has to be satisfied by the microgrid. Thus, the amount of energy that the microgrid should possess after trading energy should be greater than or equal to $F(A_{t}^{i})$.

Thus, 
\begin{align}
    & ne_{t}^{i}+d_{t}^{i}-u_{t}^{i} \geq F(A_{t}^{i}), ~ \text{which implies} \\ 
    & ne_{t}^{i} +d_{t}^{i}-F(A_{t}^{i}) \geq u_{t}^{i}.
\end{align}

After the transactions are completed, the excess energy that remains is stored in the battery for future use. The battery state  ($b_{t}^{i}$) is updated as follows:
\begin{align}
    b_{t+1}^i= \max(0,ne_{t}^{i}-u_{t}^{i}-F(A_{t}^{i})).
\end{align}


\section{Simulation Setup}
In this section, we describe the simulation setups for our experiments and appropriate models used for comparison purposes. The microgrids used in the experiments either use wind or solar renewable energies as their source. In order to simulate the renewable energy generation for all our experiments, we use the RAPsim software \cite{rapsim}. 

For comparison purposes, we also implement the constant pricing model described below:

\begin{itemize}
\item{\textbf{Constant Price Model:} The microgrids considered in this case sell energy at the constant grid price decided by the central grid. However, as highlighted in the transaction constraints, the energy is sold to the central grid at a price of $gp-k$. Please note that this model is currently being utilised in some of the power markets where the price of the transaction is decided entirely by the central grid.}
\end{itemize}

We implement our proposed dynamic pricing model and constant pricing model on following three setups:

\begin{enumerate}
\item{\textbf{Setup 1:} We first consider a simple three-microgrid setup where two of them operate on solar while the third microgrid operates on the wind renewable source. Moreover, two microgrids adopt the proposed dynamic pricing scheme while the third microgrid employs constant pricing scheme. The objective of this setup is to understand the dynamics of energy transactions between the three agents and to demonstrate the advantage of dynamic pricing over constant pricing.}
\item{\textbf{Setup 2:} Next we consider a more practical setup with 8 microgrids - four generate energy via solar farms and four generate energy via wind farms. In this setup, all microgrids generate less power than their demand at most times. We have run this setup under both models - the proposed dynamic pricing model and the constant pricing model.}
\item{\textbf{Setup 3:} This setup is similar to setup 2 with the main difference being the fact that the total renewable energy generated by the microgrids is generally more than  the total renewable energy generated by the microgrids in setup 2 while keeping demands the same.  This is to ensure that microgrids have more energy to sell as compared to setup 2. We consider eight microgrids, two of them operating on solar and six of them on wind renewable source. Such a configuration is considered to simulate the case were the majority of the microgrids generate higher electricity than the microgrids of setup 2 without violating the stability constraints.}
\end{enumerate}

\begin{figure*}[t]
    \includegraphics[scale = 0.18]{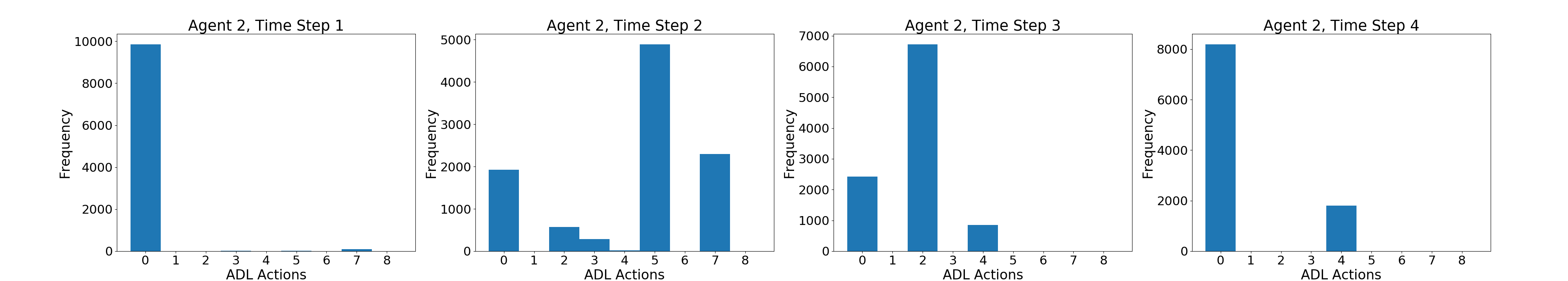}
    \caption{ADL job scheduling by the Dynamic Pricing Model under Setup 1}
    \label{f2}
\end{figure*}

\begin{figure*}[t]
    \includegraphics[scale = 0.16]{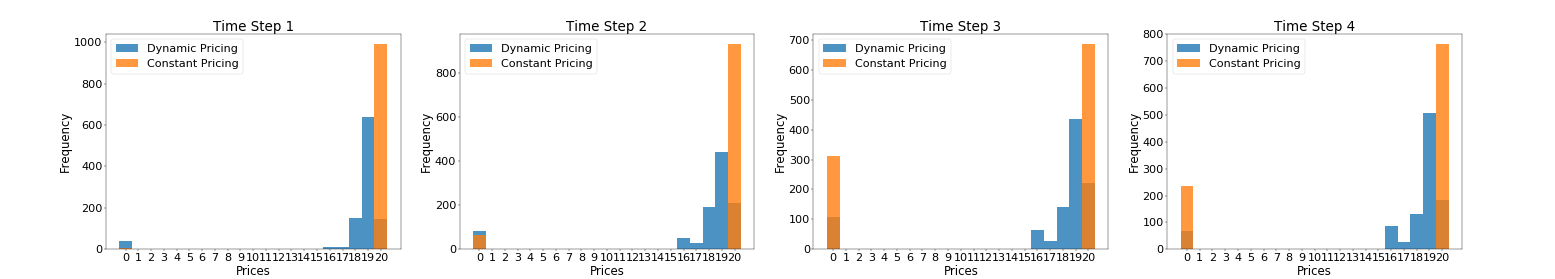}
    \caption{Pricing comparision between the Dynamic Pricing Model and the Constant Pricing Model under Setup 1}
    \label{f3}
\end{figure*}

As mentioned above, we use RAPsim simulator to generate the necessary per hour renewable energy data for all the microgrids. We then fit a Poisson distribution on this data and sample renewable energy units from this distribution during our experiments. We limit the maximum amount of electricity that can be generated from renewable sources to $10$ units and consider four decision time intervals in each day. At each epoch, the non-ADL demand can be one of the four units: $3$, $4$, $5$ or $6$. We consider three ADL jobs at the start of the day\footnote[4]{Please note that we have also carried out experiments under the stochastic ADL setting  where ADL demands appear in stochastic fashion. Please refer to the \url{https://github.com/marl-smart-grids/energy-trading/blob/master/Supplementary.pdf} for additional results.}. The maximum amount of energy that can be stored in the battery is limited to $10$ units. Similarly, the maximum amount of energy that can be bought from other microgrids or the central grid in a single time period is also limited to $10$ units. We consider a constant central grid price of $gp = 20$ (price unit per electricity unit) for our experiments. Recall that, in order to ensure cooperation among microgrids, the selling price to the central grid is fixed at $gp - k$. In our experiments, we set the value of $k$ to be 5. Therefore, the action space for the dynamic pricing strategy is $[15,20]$ price units. Both the ADL and ET agents of the microgrid use a feed forward neural network model with three layers. The complete configuration of all the microgrids along with the detailed description of the neural network and the code for all our experiments is available at the anonymous GitHub link: \url{https://github.com/marl-smart-grids/energy-trading}.

\section{Results and Discussions}
In this section, we discuss the results of our experiments. We present and discuss them under three setups as defined above. 
\begin{table*}[t]
\renewcommand{\arraystretch}{1.3}
\begin{tabular}{|l|c|c|c|c|}
\hline
\multirow{2}{*}{\textbf{Microgrid}} & \multicolumn{2}{c|}{\textbf{Rewards Obtained By Following:}}                                                             & \multirow{2}{*}{\textbf{Difference in Rewards}} & \multirow{2}{*}{\textbf{Winning Policy}} \\ \cline{2-3}
                                    & \multicolumn{1}{l|}{\textbf{Dynamic Pricing Policy (DPP)}} & \multicolumn{1}{l|}{\textbf{Constant Pricing Policy (CPP)}} &                                                 &                                          \\ \hline
1                                   & -13.094589                                                 & -13.099846                                                  & 0.005257                                        & DPP                                      \\ \hline
2                                   & -16.335466                                                 & -16.901891                                                  & 0.566425                                        & DPP                                      \\ \hline
3                                   & -37.476378                                                 & -37.532283                                                  & 0.055905                                        & DPP                                      \\ \hline
4                                   & -54.792268                                                 & -54.914326                                                  & 0.122058                                        & DPP                                      \\ \hline
5                                   & -47.218179                                                 & -47.913392                                                  & 0.695213                                        & DPP                                      \\ \hline
6                                   & -38.744498                                                 & -39.517351                                                  & 0.772853                                        & DPP                                      \\ \hline
7                                   & -55.353694                                                 & -54.785216                                                  & -0.568478                                       & CPP                                      \\ \hline
8                                   & -64.276123                                                 & -63.844191                                                  & -0.431932                                       & CPP                                      \\ \hline
\end{tabular}
\caption{Average rewards obtained by the Dynamic and Constant pricing models under setup 2}
\label{t2}
\end{table*}

\begin{table*}[]
\renewcommand{\arraystretch}{1.3}
\begin{tabular}{@{}|c|c|c|c|c|@{}}
\hline
\multirow{2}{*}{\textbf{Microgrid}} & \multicolumn{2}{c|}{\textbf{Rewards Obtained By Following:}}                                                             & \multicolumn{1}{r|}{\multirow{2}{*}{\textbf{Difference in Rewards}}} & \multicolumn{1}{l|}{\multirow{2}{*}{\textbf{Winning Policy}}} \\ \cline{2-3}
                                    & \multicolumn{1}{l|}{\textbf{Dynamic Pricing Policy (DPP)}} & \multicolumn{1}{l|}{\textbf{Constant Pricing Policy (CPP)}} & \multicolumn{1}{r|}{}                                                & \multicolumn{1}{l|}{}                                         \\ \hline
1                                   & -8.83435                                                   & -10.03937                                                   & 1.20502                                                              & DPP                                                           \\ \hline
2                                   & -1.47279                                                   & -1.03270                                                    & -0.44006                                                             & CPP                                                           \\ \hline
3                                   & 9.11136                                                    & 9.75535                                                     & -0.64399                                                             & CPP                                                           \\ \hline
4                                   & 14.65479                                                   & 14.26748                                                    & 0.38731                                                              & DPP                                                           \\ \hline
5                                   & -37.28861                                                  & -37.72918                                                   & 0.44057                                                              & DPP                                                           \\ \hline
6                                   & -47.02739                                                  & -47.52489                                                   & 0.49750                                                              & DPP                                                           \\ \hline
7                                   & -39.48666                                                  & -42.19527                                                   & 2.70861                                                              & DPP                                                           \\ \hline
8                                   & -16.90946                                                  & -17.86470                                                   & 0.95524                                                              & DPP                                                           \\ \hline
\end{tabular}
\caption{Average rewards obtained by the Dynamic and Constant pricing models under setup 3}
\label{t1}
\end{table*}

\begin{figure}
    \centering
    \includegraphics[scale = 0.36]{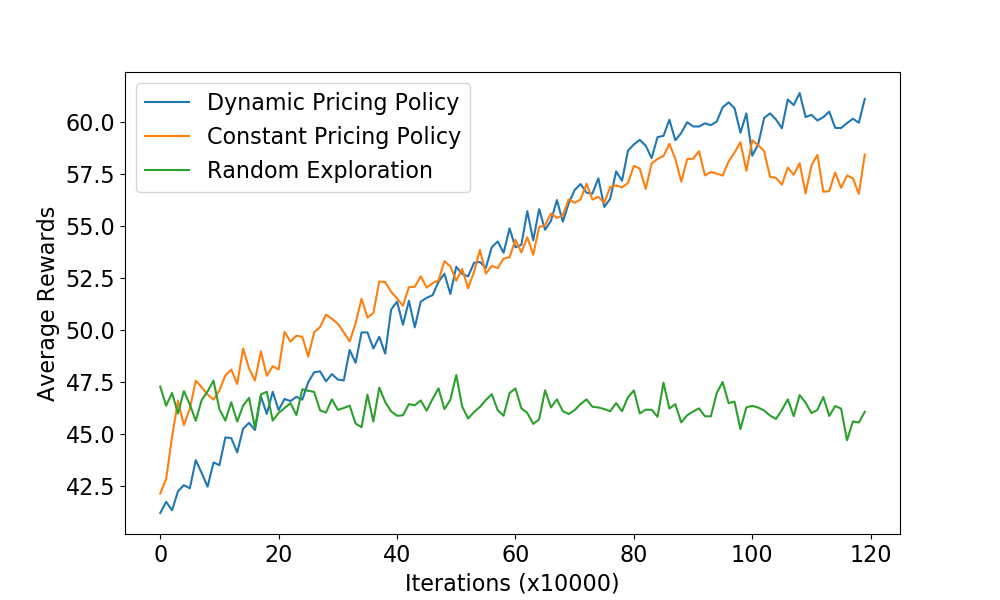}
    \caption{Comparison of all the three models with respect to average rewards under Setup 1}
    \label{f1}
\end{figure}

\textbf{Setup 1:} From the results of the experiments \footnote{Please note that due to space constraints we could not include all the results of the experiments. However, the same is available at \url{https://github.com/marl-smart-grids/energy-trading/blob/master/Supplementary.pdf}}, we make the following conclusions:
\begin{enumerate}
    \item All the 3 microgrids converge to a policy that gives higher rewards than random exploration. 
    \item We can see from Figure \ref{f1} that the agent which follows dynamic pricing (let us call it dynamic grid) obtains a higher profit than the microgrid which sells at a constant price (constant grid) when there is sufficient power generation. This may be counter-intuitive at first sight for the reason that, the selling price of the energy by the dynamic grid is always lower than the price of the constant grid (which is fixed at all times).  However, the third microgrid, during the process of buying power, when presented with two options prefers to buy from the microgrid that quotes a lower price, i.e., from the dynamic grid. In this way the dynamic grid successfully sells its power to the third microgrid at most times. The constant grid is left with no choice but to sell to the central grid at a much lower selling price ($GP-k$), yielding lower profits to it.

    \item We can observe from Figure \ref{f2} that the agents learn to schedule the ADL demands at different times which shows that our model is capable of shifting power consumption from the peak demand time. We can also observe that the ADL agent picks a certain ADL action frequently for different time steps which show the convergence of the ADL agent’s policy. In the figure, frequency denotes the number of times an ADL demand is fulfilled for the last 10,000 iterations (after convergence), at that particular time step. 
    \item We can see from Figure \ref{f3}, the dynamic nature of the prices decided by the microgrid at different times. These prices are dependent on the current state of the microgrid and are decided at each time period. In the figure, frequency denotes the number of times a particular price is selected, for the last 10,000 iterations (after convergence). From this figure, we can observe that the dynamic grid has learned to quote a price of 19 for the majority of times. This would imply that the dynamic grid has learned to adapt to the constant grid and quote a price lower than the one quoted by the constant grid which is 20. Therefore, the dynamic grid successfully sells its power to the third microgrid leading to more profits.
\end{enumerate}

\textbf{Setup 2:} From the results of our experiments, we draw the following observations:
\begin{enumerate}
    \item In Table \ref{t2}, we report the average rewards obtained by the dynamic pricing and constant pricing policies over the last $50,000$ iterations (after convergence). 
    \item From Table \ref{t2}, we observe that the proposed dynamic pricing model performs better the constant pricing model for the majority (six out of eight) of microgrids. 
    \item This empirically shows that following a DPP as compared to a CPP proves to be more advantageous to a majority of the microgrids, when energy trading occurs between microgrids.
\end{enumerate}

\textbf{Setup 3:} Through the results of our experiments, we draw the following observations:
\begin{enumerate}
    \item From Table \ref{t1}, we observe that most of the microgrids receive higher rewards through dynamic pricing than they would have accumulated through constant pricing. Therefore, our proposed dynamic pricing model proves to be more advantageous than the constant pricing model for the majority of the microgrids. 
    
    It can be observed that the microgrids achieve better rewards in setup 3 than in setup 2 (difference in rewards are higher in setup 3 compared to setup 2). We attribute this to the fact that the majority of the microgrids have higher energy generation in setup 3 as compared to setup 2, which enables them to sell more energy. Moreover, the effect of dynamic pricing becomes more prominent when they start generating more power, as noticed in the difference between their dynamic pricing rewards and constant pricing reward The differences (column 4 of Tables 1 and 2) are observed to be more in favour of dynamic pricing in setup 3 as our proposed model enables each microgrid to quote prices judiciously throughout the day, enabling them to sell intelligently and the more the energy they possess, the more does this intelligent selling reflect positively in their overall reward.
\end{enumerate}

\begin{remark}
The objective of our experiments is to understand the behavior of the microgrids learning their strategies together in a network. We wanted our models to be as close to real scenarios as possible where all the microgrids learn their policies parallelly.
\end{remark}
\begin{remark}
It is to be noted that in setup 1, we had two microgrids that generally generated more renewable energy than their demands. This was done to understand the dynamics of energy transactions between the agents. Since all the demands were satisfied and the microgrids were able to sell energy, the rewards were positive. In setup 2 and 3 we implemented more realistic scenarios (where most microgrids generally generated less energy than their demand) where experiments were performed to compare the performance of microgrid networks following dynamic pricing and constant pricing policies. Under these setups, the microgrids may have to buy energy from other grids or the main grid to fulfill some of their demands which in turn ends up creating a negative reward.
\end{remark}
From these three setups, it is clear that the agents which follow our dynamic pricing strategy are generally performing better than the constant pricing model. Moreover, we have also shown that besides dynamic pricing the microgrids also learn to intelligently schedule the ADL demands in a way that shifts the energy consumption away from peak demand.
\section{Conclusion}
In this work, we have constructed a stochastic game framework involving a network of microgrids that enables the energy trading, dynamic pricing and job scheduling. In order to solve this problem, We have devised a novel two network model (ET and ADL networks) that performs both dynamic pricing and demand scheduling at the same time. To compute the optimal policies under various setups, we have applied our proposed algorithm and have shown that the rewards obtained by our proposed dynamic pricing models yield greater rewards to the majority of the microgrids. We believe that such a modelling scheme can be applied to other sequential learning tasks.

As a future work, we would like to introduce an auction mechanism to enable transactions between microgrids where buyer microgrids can negotiate the prices decided by the seller microgrids.


\bibliographystyle{IEEEtran}
\bibliography{references}  

\end{document}